\documentclass[aps,pra,twocolumn,showpacs,amsmath,superscriptaddress]{revtex4}
\usepackage{epsfig}
\usepackage{color}

\newcommand{\ful}{\mbox{C$_{60}$}}
\newcommand{\fuls}{\mbox{\scriptsize C$_{60}$}}

\newcommand{\vscf}{\mbox{$\delta V$}}
\newcommand{\eq}[1]{Eq.~(\ref{#1})}
\begin{document}
\preprint{}
\title{Photoionization of bonding and antibonding-type atom-fullerene hybrid states in Cd@$\ful$ vs Zn@$\ful$}

\author{Mohammad H. Javani}
\affiliation{Department of Physics and Astronomy, Georgia State University, Atlanta, Georgia 30303, USA}

\author{Ruma De}
\affiliation{Center for Innovation and Entrepreneurship, Department of Natural Sciences, Northwest Missouri State University,
Maryville, Missouri 64468, USA}

\author{Mohamed E. Madjet}
\affiliation{Qatar Energy and Environment Research Institute (QEERI), P.O Box 5825, Doha, Qatar}

\author{Steven T. Manson}
\affiliation{Department of Physics and Astronomy, Georgia State University, Atlanta, Georgia 30303, USA}

\author{Himadri S. Chakraborty}
\email{himadri@nwmissouri.edu}
\affiliation{Center for Innovation and Entrepreneurship, Department of Natural Sciences, Northwest Missouri State University,
Maryville, Missouri 64468, USA}

\date{\today}

\begin{abstract}

Powerful hybridization of the Cd 4$d$ state with the $d$-angular momentum state of $\ful$ $\pi$ symmetry is found in the local density 
approximation (LDA) structure of Cd@$\ful$ ground state. The photoionization of the resulting symmetric 
and antisymmetric levels are computed using the time dependent LDA method to include electron correlations. Cross sections 
exhibit effects of the $\ful$ plasmonic motion coherently coupled to the diffraction-type cavity oscillations induced by local emissions 
from $\ful$. The Cd@$\ful$ results exhibit a substantial difference from our previous results for Zn@$\ful$. 

\end{abstract}


\maketitle

\section{Introduction}
Photoionization of atoms caged in fullerenes is an active field of current research \cite{dunsch07,popov2013}. These endofullerene materials 
have a broad scope of applications from quantum computation \cite{harneit07} to photovoltaics \cite{ross09} to drug delivery \cite{melanko09}. 
Also, endofullerenes are excellent natural laboratories to develop fundamental insights into the spectroscopy of atoms in confinement, 
as well as of doped fullerenes. Some success in the synthesis of these materials has spawned inspiring recent 
experiments \cite{scully05,mueller08,kilcoyne10,phaneuf13}. 

Our jellium-based time-dependent local density functional technique \cite{madjet10} is one of the most complete approach among various theoretical models 
of photoionization studies \cite{lo09,amusia-jpb-08,dolmatov08,govil09,stener02,chen12,gorczyca12,jose13}. Because the scheme incorporates the ground state, excited states and 
dynamical interactions of all atomic-electrons and active fullerene electrons in the same computational footing, 
including significant aspects of correlation. 
Using this method, we have already found (i) a strong enhancement in the atomic photoresponse from the host fullerene's plasmon 
dynamics \cite{madjet07,chakraborty08,javani12} and (ii) atom-fullerene hybrid levels with novel ionization behavior \cite{chakraborty09,madjet10}. 
The latter feature is quite interesting, since these hybrid levels lead to possibilities of covalent-type bonding of the fullerene with the 
trapped atom and their influence on the ionization response of the compound to electromagnetic radiation or charged-particles impact. In general, 
spectroscopic examination of such hybrids pave the route to probe wavefunction mixing in other spherical dimer composites, such as,
buckyonions or clusters trapped in fullerene cages.

Only the atomic and fullerene orbitals of the same angular momentum hybridize. This is because the orthogonality property of the spherical harmonics 
makes the other terms zero. Also, from a perturbation theory viewpoint, to have strong hybridization not only good overlap of the unperturbed 
wave functions is needed but the binding energies of those levels have to be close, since they respectively guarantee a large numerator and and a small 
denominator of the coupling term. These conditions applied to the known energy levels of the $\ful$ $\pi$ band \cite{madjet-jpb-08} 
suggest that the atom's valence and subvalence levels are susceptible to hybridization.  For Xe@$\ful$ a strong $s$-$s$ hybridization between 
Xe 5$s$ and a $\ful$ s level of $\pi$ character was predicted \cite{chakraborty09}. Also in the case of Zn@$\ful$ strong $d$-$d$ hybridization have been found 
between Zn 3$d$ and a $\ful$ $d$ level \cite{maser12}. Here we investigate the same phenomenon in Cd@$\ful$ to explore how much
the detail shapes of its hybrid wavefunctions alter from the corresponding Zn@$\ful$ hybrids given an extra node in 
Cd 4$d$.  The photoionization cross sections 
of these hybrid levels exhibiting structures over a broad energy range from effects of plasmons and oscillatory modulations in the emission process are calculated. 
In addition, the results are compared and contrasted with the earlier Zn@$\ful$ results to examine the extent of their differences 
from differing emission responses of Zn and Cd.

\section{Ground-state hybridization} 

We employ nonrelativistic density functional theory to describe the electronic structure of the $\ful$ cage; the details of the method can be found 
in Ref.\,[\onlinecite{madjet-jpb-08}]. Previous works (i) explained the measured oscillations in the valence photoelectron intensities 
of neutral $\ful$ \cite{ruedel02} and (ii) agreed with an experimental study of plasmons by predicting a new high-energy 
plasmon resonance in the photoionization of $\ful$ cations \cite{scully05}. In the formulation of the $\ful$ ground state, the four 
valence electrons ($2s^2$$2p^2$) of each carbon atom are allowed to delocalize. But the core of C$^{4+}$ ions (each consisting of 
a carbon nucleus plus two tightly-bound 1s electrons) are represented by a spherical jellium shell with a radius $R$=3.54\AA\, and a thickness $\Delta$, 
{\em plus} an adjustable pseudo-potential $V_0$ \cite{puska93}. An earlier Hartree-Fock study using a molecular orbital basis for Zn@$\ful$
indicated a weak Zn-$\ful$ ionic bond with a central position of Zn in the cage and that the compound remains stable up to the degradation-temperature
of the $\ful$ frame \cite{varganov00} . Encouraged, we assume similar central position of Cd, another Group IIB metal, in the sphere. Then the Kohn-Sham 
equations for the system of 288 electrons (48 from Cd and 240 delocalized electrons from the $\ful$ cage) are solved to obtain the 
ground state in the local density approximation (LDA). A widely used parametric formulation is employed \cite{gunnerson76} to approximately treat the electrons' 
exchange-correlation interactions. $V_0$ and $\Delta$ are determined by requiring charge neutrality and by producing the experimental value, 
$-$7.54 eV, of the first ionization potential. The width $\Delta$ is found to be 1.5\AA, which agrees well with experimentally derived value of 
the molecular width \cite{ruedel02}. Further, although the jellium ignores the truncated-icosahedral C-structure, similarities of our LDA 
ground state near the HOMO and HOMO-1 levels with known quantum chemical calculations \cite{troullier92} were noted in Ref.\,[\onlinecite{madjet-jpb-08}].
\begin{figure} 
\includegraphics[height=10.0cm,width=8.5cm,angle=0]{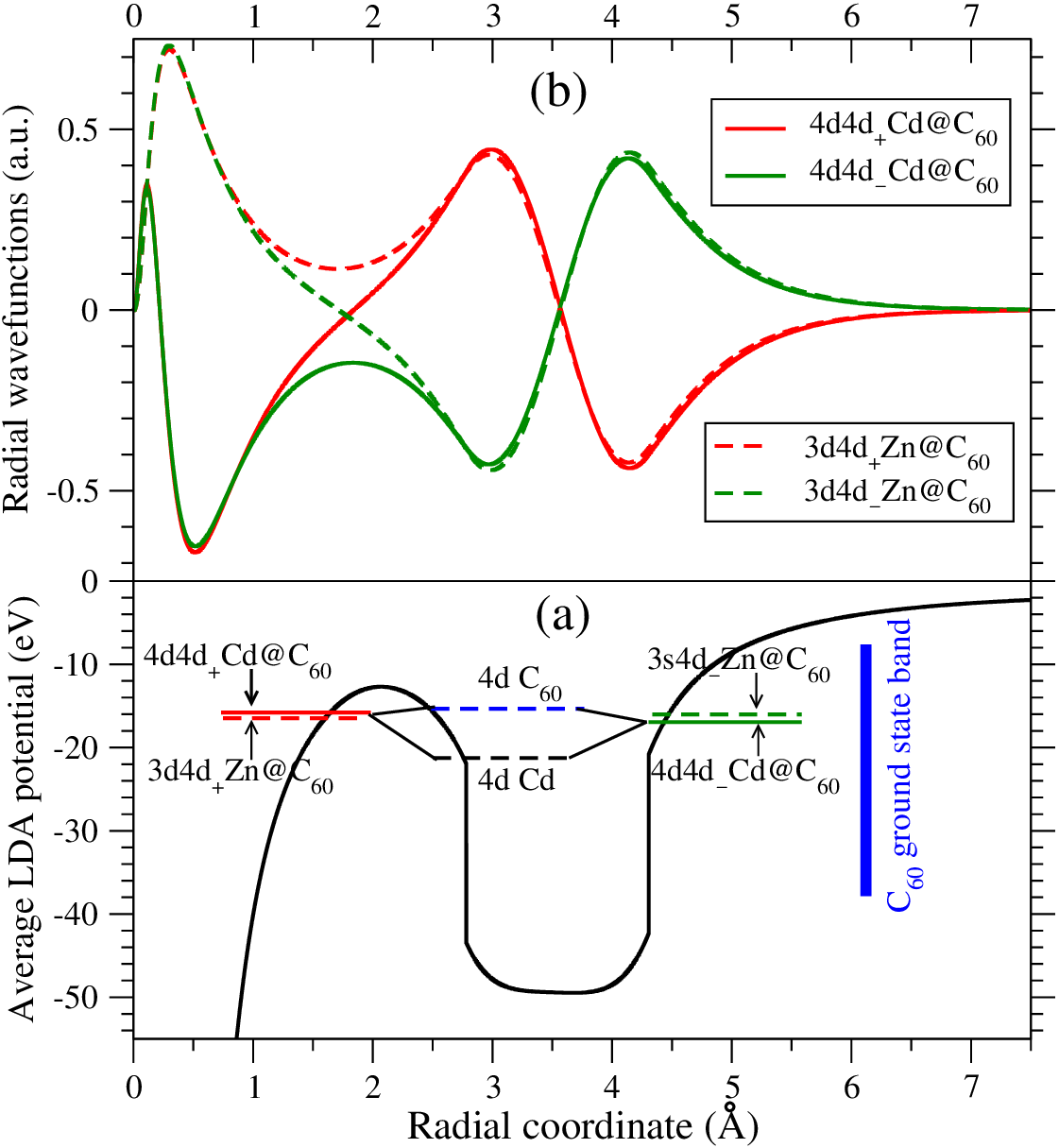} 
	\caption{(Color online) (a) The average radial LDA ground state potential of Cd@$\ful$. $\ful$ single-electron occupied band, 4$d$ Cd and 4$d$ $\ful$ 
	unperturbed levels, and two Cd@$\ful$ hybrid levels are shown. Similar hybrids in Zn@$\ful$ are included for comparison.
	(b) Four radial wavefunctions for bonding and antibonding hybrid states of both Cd@$\ful$ and Zn@$\ful$.}
	\label{fig:cd@c60-figure1}
\end{figure}

The asymptotic behavior of the LDA ground-state potential was improved by employing a self-interaction correction scheme as done by Perdew and Zunger \cite{perdew81}, 
along with a similar implementation for the excited states \cite{saito91}. Consequently, this correction results in orbital-specific single electron 
potentials. Thus, the {\em effective} radial potential, averaged over the orbitals and weighted by their occupancies, is presented in Fig.\,1(a). A powerful 
hybridization between the free (unperturbed) Cd 4$d$ state and $\ful$ 4$d$ ($\pi$) state occurs. (Standard Coulomb notation is used to label both Zn and $\ful$ orbitals.) 
States hybridized in any arbitrary proportion of two constituent states $|\phi_{4d {\mbox{\scriptsize Cd}}}\rangle$ and $|\phi_{4d \fuls}\rangle$ of free systems 
can be described as symmetric and antisymmetric combinations:
\begin{subequations}\label{bound-hyb}
\begin{equation}\label{bound-hib1}
4d4d_+ = |\phi_+\rangle = \sqrt{\alpha}|\phi_{4d \mbox{\scriptsize Cd}}\rangle + \sqrt{1-\alpha}|\phi_{4d \fuls}\rangle
\end{equation}
\begin{equation}\label{bound-hib2}
4d4d_- = |\phi_-\rangle = \sqrt{1-\alpha}|\phi_{4d \mbox{\scriptsize Cd}}\rangle - \sqrt{\alpha}|\phi_{4d \fuls}\rangle,
\end{equation}
\end{subequations}  
that embody the bonding and antibonding Cd-$\ful$ interactions; to preserve normalization $\alpha$ must be between zero and one. Energies of the two 
reactant-levels of pristine systems and their hybrid products are shown in Fig.\,1(a), and the radial hybrid wavefunctions in Fig.\,1(b). It is found that multiplying the $4d4d_+$ 
wavefunction by a factor of $\sqrt{2}$ reproduces the Cd 4$d$ and $\ful$ 4$d$ wavefunctions (not shown) in their respective regions, i.e., the hybridized states are roughly a 
50-50 admixture of the electron densities so that $\alpha$ in Eqs.\,(\ref{bound-hyb}) is about $\frac{1}{2}$. It is surprising that so strong a hybridization 
occurs even though the overlap between Cd 4$d$ and $\ful$ 4$d$ wavefunctions (not shown) is found small, as in Zn@$\ful$ \cite{maser12}; thus, it must be due to the 
near-degeneracy of the participant 
levels [Fig.\,1(a)]. This property puts these hybrid states, along with those predicted earlier \cite{chakraborty09,maser12}, in a different league than 
the known hybrids \cite{jackson94} from physical contacts of a larger atom packed in a smaller fullerene.

Radial wavefunctions that emerge from Zn 3$d$ hybridized with $\ful$ 4$d$ in the Zn@$\ful$ compound is also displayed in Fig.\,1(b). Comparison reveals differences
between wavefunctions of the two compounds for each of bonding and antibonding symmetries, which is owing to the node that Cd 4$d$ has whereas Zn 3$d$ is nodeless.
In fact, this distinction is also the reason why $4d4d_+$ is less bound than $4d4d_-$ in Cd@$\ful$ while the trend is just the opposite in Zn@$\ful$ [Fig.\,1(a)]; in 
each case, the binding energy of the hybridized state with the larger number of nodes is decreased, as they must.

\section{Photoionization: brief theory}

By using the time-dependent LDA (TDLDA) methodology \cite{madjet-jpb-08}, the 
response of the system to the external field is obtained. The perturbation $z$, the dipole 
interaction for linearly polarized light, induces a frequency-dependent complex change in the electron 
density arising from dynamical electron correlations. This can be written, using the LDA susceptibility $\chi_0$, as
\begin{equation}\label{ind_den2}
\delta\rho({\bf r};\omega)={\int \chi_0({\bf r},{\bf r^{\prime}};\omega) 
                           \vscf({\bf r^{\prime}};\omega) d{\bf r^{\prime}}},
\end{equation}
in which
\begin{equation}\label{v_scf}
\vscf({\bf r^{\prime}};\omega) = z +\!\int\!\!\frac{\delta\rho({\bf r^{\prime}};\omega)}
     {\left|{\bf r}-{\bf r^{\prime}}\right|}d{\bf r^{\prime}}
     \!+\!\left[\frac{\partial V_{\mbox{xc}}}{\partial \rho}\right]
     _{\rho=\rho_{0}}
     \!\!\!\!\delta\rho({\bf r};\omega),
\end{equation}
where the second and third term on the right hand side are, respectively, the induced change of the Coulomb and the 
exchange-correlation potentials. Clearly, besides containing the perturbation $z$, $\vscf$ 
also includes the dynamical field produced by important correlations.
The photoionization cross section is then obtained as the sum of independent channel cross sections
$\sigma_{n\ell\rightarrow k\ell'}$, corresponding to a dipole transition $n\ell\rightarrow k\ell^\prime$:
\begin{equation}\label{cross_pi}
\sigma_{\mbox{\scriptsize{PI}}}(\omega)\!=\!\!\sum_{n\ell}\!\!\sigma_{n\ell\rightarrow k\ell'}
    \sim\!\! \sum_{n\ell}\!2(2\ell+1)|\langle \phi_{k\ell'}|\vscf|\phi_{n\ell}\rangle|^2.
\end{equation}
Note that, replacing $\vscf$ in Eq.\ (\ref{cross_pi}) by $z$ yields the LDA cross section that entirely omits the correlation. 
\begin{figure}
\includegraphics[height=7.cm,width=8.cm,angle=0]{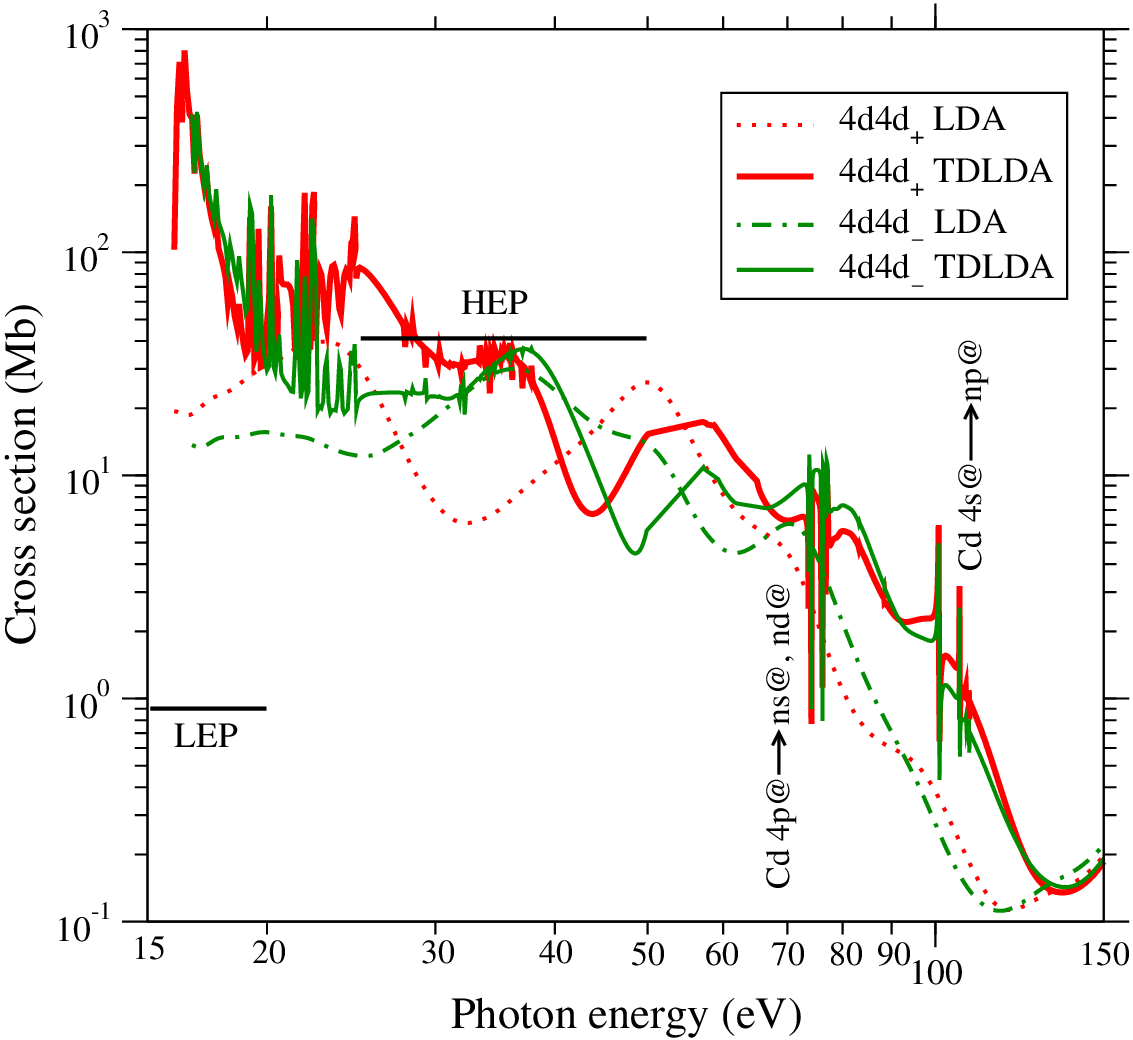} 
	\caption{(Color online) Photoionization cross sections by single-electron LDA and  
many-electron TDLDA methods for Cd@$\ful$ hybrid states. Autoionizing resonances for inner shell excitations from $\ful$, appearing at lower photon energies, 
and from 4$p$@ and 4$s$@ states of Cd are identified. $\ful$ low (LEP) and high energy plasmon (HEP) regions are indicated.}
	\label{fig:cd@c60-figure2}
\end{figure}

\section{Results and discussion}

The single-electron LDA photoionization cross sections, as a function of the photon energy, for the two hybrid states $4d4d_{\pm}$ are presented in 
Fig.\,2. They are seen to be substantially different, both in magnitudes and structures, from the LDA cross sections of Cd 4$d$ and $\ful$ 4$d$ levels in Fig.\,3.  
This is the effect of the wavefunction mixing merely via hybridization as the correlation is omitted in LDA. The mechanism of the oscillatory structure
can be described by the {\em acceleration} gauge form of the dipole matrix element which will be discussed below.
\begin{figure}
\includegraphics[height=7.5cm,width=8.5cm,angle=0]{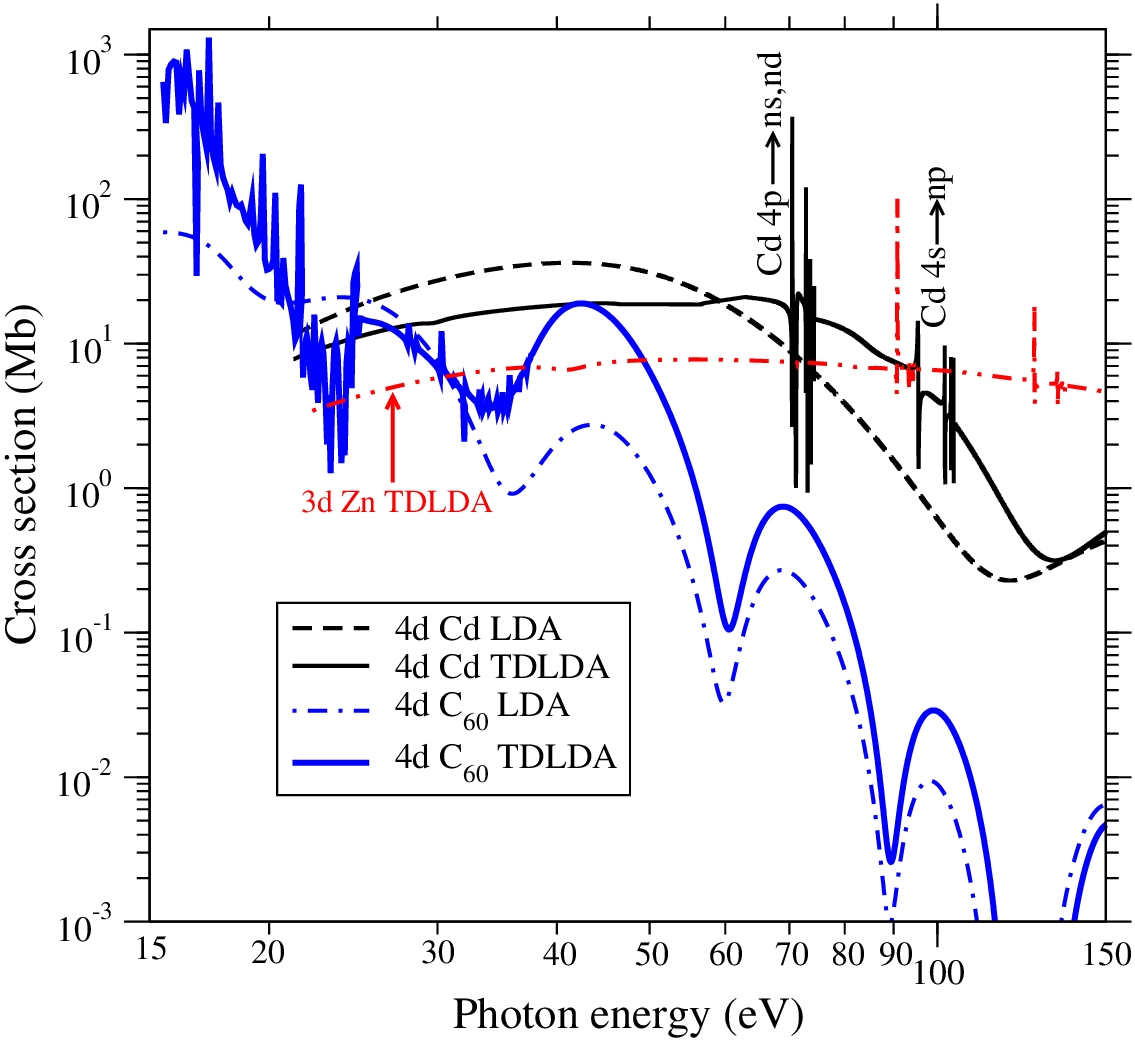} 
	\caption{(Color online) LDA and TDLDA cross sections for 4$d$ of free Cd and 4$d$ of empty $\ful$. Autoionizing resonances due to Cd 4$p$ excitations are identified.
	The TDLDA result for 3$d$ level of free Zn is included for comparison.}
	\label{fig:cd@c60-figure3}
\end{figure}

Cross sections of the hybrid levels, obtained using TDLDA and shown in Fig.\,2, should be rather realistic quantitatively, since they include 
correlation effects. The TDLDA cross sections exhibit $\ful$ autoionizing resonances at low energies, along with Cd 4$p@ \rightarrow ns@,nd@$, 4$s@ \rightarrow np@$ 
autoionizing resonances at higher energies; the symbol $n\ell$@ is used to denote the levels of the confined atom and @$n\ell$ to represent the levels of the doped 
$\ful$. These TDLDA results in Fig.\,2 dramatically modify the LDA cross sections in the fullerene's low energy plasmon (LEP) region centered around 17 eV (Fig.\,3), and also 
over the broader region of the high energy plasmon (HEP), showing substantial differences as high as up to 40 eV. Obviously, in the TDLDA results for hybrid 
ionization, correlation-driven many-electron plasmonic enhancement from $\ful$ couples with the single-electron oscillation effects seen in LDA. The following 
diagnostics are useful to better interpret the results.

The general correlation-modified (TDLDA) matrix element, in the framework of the first-order perturbation theory based interchannel coupling, of the dipole 
photoionization of 4$d$4$d_\pm$ levels can be written as \cite{javani12},
\begin{eqnarray}\label{gen-mat-element}
{\cal M}_\pm (E) &=& {\cal D}_\pm (E)\nonumber \\
                 &+& \!\!\!\!\!\displaystyle\sum_{n\ell}\!\!\!\int\!\!\! dE' \!\frac{\langle\psi_{n\ell}(E')|\frac{1}{|{\bf r}_{\pm}-{\bf r}_{n\ell}|}
|\psi_{\pm}(E)\rangle}{E-E'} \!{\cal D}_{n\ell} (E')
\end{eqnarray}
where the single electron (LDA) bound-continuum matrix element
\begin{equation}\label{se-mat-element}
{\cal D}_\pm (E) = \langle kp(f)|z|\phi_\pm\rangle
\end{equation}
and $|\psi\rangle$ is the $n\ell\rightarrow k\ell'$ {\em channel} wavefunction.

Taking the hybridization into account, the channel wavefunctions in \eq{gen-mat-element} become
\begin{subequations}\label{channel-hyb}
\begin{equation}\label{channel-hib1}
|\psi_+\rangle = \sqrt{\alpha}|\psi_{4d@ \mbox{\scriptsize Cd}}\rangle + \sqrt{1-\alpha}|\psi_{@4d \fuls}\rangle
\end{equation}
\begin{equation}\label{channel-hib2}
|\psi_-\rangle = \sqrt{1-\alpha}|\psi_{4d@ \mbox{\scriptsize Cd}}\rangle - \sqrt{\alpha}|\psi_{@4d \fuls}\rangle.
\end{equation}
\end{subequations}  
In Eqs.\,(\ref{channel-hyb}) the subscripts 4$d$@ and @4$d$ are used to include the modifications of the continuum waves of, respectively, the confined Cd and the doped $\ful$.

Substituting Eqs.\ (\ref{bound-hyb}) and (\ref{channel-hyb}) in \eq{gen-mat-element}, and noting that the overlap between a pure Cd and a pure $\ful$ bound 
state is negligible, we can easily separate the atomic and fullerene contributions to the integral to get the TDLDA matrix element for the symmetric hybrid $4d4d_+$ as
\begin{widetext}
\begin{eqnarray}\label{gen-mat-element+}
{\cal M}_+ (E) &=& \sqrt{\alpha}\left[{\cal D}_{4d@ \mbox{\scriptsize Cd}} (E)
                 + \displaystyle\sum_{n\ell (\mbox{\scriptsize Cd})}\int dE' \frac{\langle\psi_{n\ell}(E')|\frac{1}{|{\bf r}_{+}-{\bf r}_{n\ell}|}
|\psi_{4d@ \mbox{\scriptsize Cd}}(E)\rangle}{E-E'} {\cal D}_{n\ell} (E')\right ] \nonumber \\
               && + \sqrt{1-\alpha}\left[{\cal D}_{@4d \fuls} (E)
                 + \displaystyle\sum_{n\ell (\fuls)}\int dE' \frac{\langle\psi_{n\ell}(E')|\frac{1}{|{\bf r}_{+}-{\bf r}_{n\ell}|}
|\psi_{@4d \fuls}(E)\rangle}{E-E'} {\cal D}_{n\ell} (E')\right] \\
               &=& \sqrt{\alpha}{\cal M}_{4d@ \mbox{\scriptsize Cd}} (E) + \sqrt{1-\alpha}{\cal M}_{@4d \fuls} (E)\label{gen-mat-element+eq}. 
\end{eqnarray}
\end{widetext}
Similarly, the TDLDA matrix element for the asymmetric hybrid level is
\begin{equation}\label{gen-mat-element-}
{\cal M}_- (E) = \sqrt{1-\alpha}{\cal M}_{4d@ \mbox{\scriptsize Cd}} (E) - \sqrt{\alpha}{\cal M}_{@4d \fuls} (E).
\end{equation}

\subsection{LDA multi-path interference oscillations}

Within the LDA framework, where correlations are omitted, both the integrals on the right-hand-side of \eq{gen-mat-element+} will vanish and 
simplify Eqs.\,(\ref{gen-mat-element+eq}) and (\ref{gen-mat-element-}) to
\begin{subequations}\label{lda-mat-elements}
\begin{equation}\label{lda-mat-element+}
{\cal D}_+ (E) = \sqrt{\alpha}{\cal D}_{4d@ \mbox{\scriptsize Cd}} (E) + \sqrt{1-\alpha}{\cal D}_{@4d \fuls} (E)
\end{equation}
\begin{equation}\label{lda-mat-element-}
{\cal D}_- (E) = \sqrt{1-\alpha}{\cal D}_{4d@ \mbox{\scriptsize Cd}} (E) - \sqrt{\alpha}{\cal D}_{@4d \fuls} (E),
\end{equation}\end{subequations}
which, of course, can also be obtained directly by substituting Eqs.\,(\ref{bound-hyb}) in \eq{se-mat-element}.

In LDA the production of the confinement oscillations is easily explained in the {\em acceleration gauge} frame where the dipole matrix element, 
\eq{se-mat-element}, is expressed as,
\begin{eqnarray}\label{amp_fund}
{\cal D}_{\pm} (E) \sim \left\langle kp(f)\left|\frac{\partial V}{\partial r}\right|\phi_{\pm}\right\rangle, 
\end{eqnarray}
which underpins the idea that the electron in the potential $V(r)$ needs a force $\partial V/\partial r$ to escape. This ionizing force produced by the average 
radial potential [Fig.\,1(a)] of the compound peaks at the inner and the outer edges, $R_i$ (= 2.79 \AA) and $R_o$ (= 4.29 \AA), of the $\ful$ shell, suggesting 
strong emission from the edge regions where the potential changes rapidly. Furthermore, Fig.\,1(a) shows that a strong force also exists in the central atomic 
region where the potential continuously varies. Thus, since the hybrid wavefunctions $\phi_{\pm}$ are finite over all these force-sites [Fig.\,1(b)], photoemission will 
occur from all three regions, significantly interfering with each other through the coherence. The effect further enriches, since the part of the amplitude emanated from Cd 
reflects from the shell as a result of the modified atomic continuum due to the surrounding $\ful$ potential. 

The general structure of this LDA matrix element has been discussed previously \cite{potter10,mccune09}. Following Ref.\,\cite{mccune09},
\begin{subequations}\label{etas-amp}
\begin{eqnarray}\label{etas@-amp-atom}
{\cal D}_{4d@ \mbox{\scriptsize Cd}} &\sim& {\cal D}^{\mbox{\tiny atom}}(k) \nonumber \\
                  & + & A^{\mbox{\tiny refl}}(k)\left[e^{-ikD_o}e^{-iV_0\frac{2\Delta}{k}} - e^{-ikD_i} \!\right]
\end{eqnarray}
\begin{equation}\label{etas@-amp-shell}
{\cal D}_{@4d \fuls} \sim  A^{\mbox{\tiny shell}}(k)e^{-i\frac{V_0}{k}}\left[a_ie^{-ikR_i}- a_oe^{-ikR_o}\!\right],
\end{equation}
\end{subequations}
where the photoelectron momentum $k =\sqrt{2(E-\epsilon_{\pm})}$ in atomic units, $V_0$ is the average depth of the shell potential, and $a_i$ and $a_o$ are the values 
of $\phi_{\pm}$ at $R_i$ and $R_o$. In \eq{etas@-amp-atom}, the ${\cal D}^{\mbox{\tiny atom}}$ is the contribution from 
the atomic region and the second term on the right hand side denotes the reflection induced oscillations in momentum coordinate with frequencies $D_i$ and $D_o$, the 
inner and the outer diameter of the shell. \eq{etas@-amp-shell} represents the portion of the overlap integral from the shell region, producing two collateral emissions 
from the edges, where non-zero ionizing forces exist; as evident, these contributions oscillate in two frequencies, $R_i$ and $R_o$. The latter effect is similar 
to the {\em diffraction in momentum space} where oscillations (fringes) are connected to the fullerene radii.   

The LDA cross sections in Fig.\,2, obtained by squaring the modulus of Eqs.\,(\ref{lda-mat-elements}), hence involve interferences among atomic, reflective and 
shell ionization modes yielding oscillations. As shown earlier \cite{madjet10} by Fourier transforming the cross sections of $s$-$s$ hybrid states in Xe@$\ful$, 
$\sigma^{\mbox{\tiny LDA}}_{\pm}$ contain dominant frequencies: $D_i$, $D_o$ from the reflective and $R_i$, $R_o$ from the diffractive shell-emissions.

\subsection{Cd-$\ful$ coherence in TDLDA}

The TDLDA cross sections of the hybrid ionization are obtained by squaring the modulus of Eqs.\, (\ref{gen-mat-element+eq}) and (\ref{gen-mat-element-}):
\begin{subequations}\label{cross-tdlda}
\begin{eqnarray}\label{cross-tdlda+}
\sigma_+ & = & \alpha\sigma_{4d@ \mbox{\scriptsize Cd}} + (1-\alpha)\sigma_{@4d \fuls} \nonumber \\
         & + & \sqrt{\alpha-\alpha^2}{\cal M}_{4d@ \mbox{\scriptsize Cd}} \otimes {\cal M}_{@4d \fuls}
\end{eqnarray}
\begin{eqnarray}\label{cross-tdlda-}
\sigma_- & = & (1-\alpha)\sigma_{4d@ \mbox{\scriptsize Cd}} + \alpha\sigma_{@4d \fuls} \nonumber \\
         & - & \sqrt{\alpha-\alpha^2}{\cal M}_{4d@ \mbox{\scriptsize Cd}} \otimes {\cal M}_{@4d \fuls}
\end{eqnarray}
\end{subequations}
where ${\cal M}_p \otimes {\cal M}_q = {\cal M}_{p}^{\ast} {\cal M}_q + {\cal M}_p {\cal M}_{q}^{\ast}$, which represents dynamical interferences between Cd 
and $\ful$ TDLDA amplitudes from their coherent superpositions.

The many-electron contribution from the Cd region of the compound, the first integral on the right hand side of \eq{gen-mat-element+}, is weak. 
This is evident from the small differences between Cd 4$d$ LDA and TDLDA curves in Fig.\,3. However, both these Cd 4$d$ curves are significantly higher 
than $\ful$ 4$d$ curves at higher energies. On the other hand, the differences between LDA and TDLDA predictions for $\ful$ 4$d$ (Fig.\,3) are huge over the 
LEP and HEP regions from the giant enhancements via the second integral in \eq{gen-mat-element+}. These enhancements 
can even mask the reflection effect in \eq{etas@-amp-atom}, while far stronger diffraction oscillations, \eq{etas@-amp-shell}, 
will remain. Consequently, the large value of ${\cal M}_{@4d \fuls}$ at lower energies and that of ${\cal M}_{4d@ \mbox{\scriptsize Cd}}$ at higher energies induce very 
strong Cd-$\ful$ coherent-mixing in $4d4d_{\pm}$ TDLDA cross sections over a broad energy range as seen in Fig.\,3.
\begin{figure}
\includegraphics[height=10.0cm,width=8.5cm,angle=0]{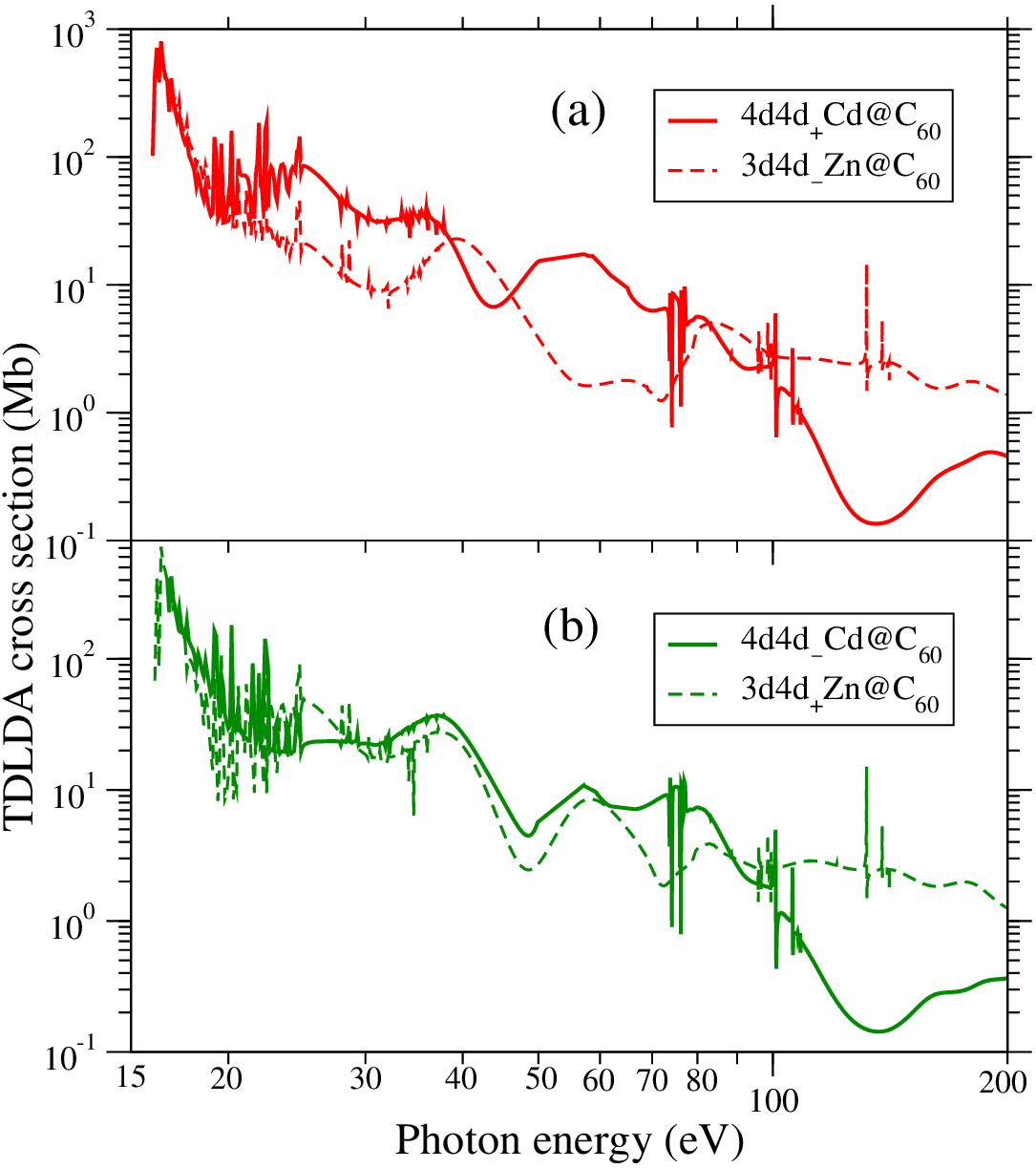} 
	\caption{(Color online) TDLDA photoionization cross sections of the bonding (upper panel) and antibonding (lower panel) hybridized $d$ states for Cd@$\ful$ and Zn@$\ful$.}
	\label{fig:cd@c60-figure4}
\end{figure}

\subsection{Cd@$\ful$ versus Zn@$\ful$}

Photoionization of hybrid levels from coupled $d$-angular momentum states has recently been studied for Zn@$\ful$ \cite{maser12}. Here we compare our
current results 
for Cd@$\ful$ with Zn@$\ful$ in Figs.\,4. Comparison reveals a number of significant differences. 
For each of symmetric and antisymmetric combinations, the $\ful$ $d$-orbital mixes with 4$d$ Cd and 3$d$ Zn by roughly the same amount [Fig.\,1(b)]. 
This allows approximately similar shapes and magnitudes of plasmon-induced enhancements in the cross sections below 20 eV in Fig.\,4. At energies above 20 eV up to 100 eV, 
the region of energy-dependent oscillations, strong disagreements between the cross sections of the bonding pair are noted. The difference is somewhat weaker 
for the antibonding pair but is still significant (note the {\em log} scale of the cross sections in Fig.\,4). But the differences in the ground state hybrid 
wavefunction structure around 2 \AA\ between Cd@$\ful$ and Zn@$\ful$ [Fig.\,1(b)] will be largely insensitive to 
their ionization behavior, since the radial range of this difference coincides with the plateau of the potential [Fig.\,1(a)] where the potential's 
derivative (ionizing force) is small. Further, \eq{etas@-amp-shell} suggests that strengths and relative phases of oscillations depend respectively on the 
magnitude and sign of hybrid wavefunctions at the shell boundaries. However, for a given hybrid level these quantities are practically equal for the two 
compounds, see Fig.\,1(b). The mismatch between the cross sections in Figs.\,4 then must be due to the differences between 4$d$ and 3$d$ emissions
of free Cd and Zn that can alter the atomic contributions in Eqs.\,(\ref{gen-mat-element+}) and (\ref{gen-mat-element-}). Indeed, as shown in Fig.\,3, significant mismatch
between TDLDA Cd 4$d$ and Zn 3$d$ curves, including a more defined shape resonance followed by a Cooper minimum in Cd at 135 eV, corroborates this assumption. 
Above 100 eV then, the remarkable differences between Cd@$\ful$ and Zn@$\ful$ 
predictions in Fig.\,4 results from the coherent mixing with 4$d$ Cooper minimum in Cd.

\section{Conclusion}

As for the summarizing remarks, a pair of bonding and antibonding orbitals in Cd@$\ful$, originating from the hybridization of outer $d$-angular momentum states of the 
free Cd atom and the empty $\ful$ molecule was found. The mixing arises from the near degeneracy of the coupling levels. These hybrid states can be seen 
as atom-shell spherical analogues of two-center dimer states known among the molecules. Photoionization from these levels shows radically different 
magnitudes and structures from those of the individual free-system states. Analysis reveals that the coherent superposition between the atomic and $\ful$ ionization 
engenders sizable effects in the hybrid emission behavior: while the lower photon-energy part of the hybrid 
cross sections are enhanced by the interaction with the shell's plasmon response, the higher energy ranges are empowered by the strength of atomic ionization, 
rendering the effect amenable to measurements over a broad window of photon energy. 

The Cd@$\ful$ results are compared to our previous results for Zn@$\ful$. They have roughly the same shape below 
20 eV, since the cross section in this region is dominated by interchannel coupling with the $\ful$ plasmon. But between 20 to 200 eV and beyond
significant disagreement is seen due to differences in their atomic emissions. Although shape differences in the hybrid wavefunctions are found in the radial zone
between the atom and $\ful$ shell, there is almost no contribution to the dipole matrix element from this region, so the differences do not appreciably affect the 
photoionization. The study suggests that there are substantial differences in the photoionization 
of outer subshells of chemically similar endohedral atoms, depending upon variations in their free atomic response.  We hypothesize that $f$-$f$ hybrids may 
exist in some lanthanide and actinide metallofullerenes, the confirmation of which is a subject of future studies.  

\begin{acknowledgments}
This work is supported by NSF and DOE, Basic Energy Sciences.
\end{acknowledgments}

\end{document}